\documentclass[12pt]{article} 

\usepackage[pdftex]{graphicx}

\usepackage{amssymb}
\usepackage{amsthm}
\usepackage{amsmath}
\usepackage[tableposition=top]{caption}
\usepackage{subcaption}
\usepackage{mathtools}
\usepackage{latexsym}
\usepackage{enumitem}
\usepackage{mleftright}
\usepackage{resizegather}
\usepackage[highlightmarkup==uwave]{changes}
\usepackage[toc,page]{appendix}
\usepackage{epstopdf}
\usepackage{float}
\usepackage{setspace}
\usepackage{soul}
\usepackage{hyperref}
\usepackage{url}
\usepackage[round, authoryear]{natbib}
\usepackage{mdframed}
\usepackage{bm}
\usepackage{multirow}
\usepackage{booktabs}
\usepackage{hyperref}

\theoremstyle{plain} 
\theoremstyle{remark}

\newtheorem{corollary}{Corollary}
\newtheorem{theorem}{Theorem}

\graphicspath{{./}{simulation-newalgrithm/}}

\newcommand{\E}{\mathrm{E}}
\newcommand{\ve}{\mathrm{vec}}
\newcommand{\vech}{\mathrm{vech}}
\newcommand{\var}{\mathrm{Var}}

\newcommand{\cov}{\mathrm{Cov}}

\newcommand{\R}{\mathbb{R}}

\newcommand{\bbeta}{\boldsymbol\beta}

\newcommand{\bGamma}{\boldsymbol\Gamma}
\newcommand{\bOmega}{\boldsymbol\Omega}

\newcommand{\bSigma}{\boldsymbol\Sigma}

\newcommand{\brho}{\boldsymbol\rho}
\newcommand{\env}{\mathcal{E}}
\newcommand{\boldeta}{\boldsymbol\eta}

\newcommand{\bzero}{\boldsymbol 0}

\newcommand{\bI}{\textbf{I}}
\newcommand{\bA}{\boldsymbol A}
\newcommand{\bB}{\textbf{B}}
\newcommand{\bC}{\textbf{C}}

\newcommand{\bZ}{\textbf{Z}}
\newcommand{\bS}{\boldsymbol S}
\newcommand{\bE}{\textbf{E}}
\newcommand{\bM}{\textbf{M}}

\newcommand{\bG}{\textbf{G}}
\newcommand{\bP}{\textbf{P}}
\newcommand{\bQ}{\textbf{Q}}

\newcommand{\bH}{\textbf{H}}

\newcommand{\bJ}{\textbf{J}}

\newcommand{\bR}{\textbf{R}}

\newcommand{\bX}{\boldsymbol X}
\newcommand{\bY}{\boldsymbol Y}
\newcommand{\tX}{\tilde{\boldsymbol X}}
\newcommand{\tY}{\tilde{\boldsymbol Y}}

\newcommand{\vY}{\vec{Y}}
\newcommand{\vy}{\vec{y}}
\newcommand{\vX}{\vec{X}}
\newcommand{\vZ}{\vec{Z}}
\newcommand{\vepsilon}{\vec{\epsilon}}

\newcommand{\vmu}{\vec{\mu}}

\newcommand{\ones}{\vec{1}}

\DeclareMathOperator*{\argmin}{\arg\!\min}

\title{Dimension Reduction for Spatially Correlated Data:  Spatial Predictor Envelope}
\author{ Paul May$^a$ and Hossein Moradi Rekabdarkolaee$^{b*}$ \\[0.2in]
\normalsize\itshape $^a$ Department of Geographical Sciences, University of Maryland,\\ 
\normalsize\itshape $^b$ Department of Mathematics and Statistics, \\
\normalsize\itshape South Dakota State University, Brookings, South Dakota.}

\date{} 

\begin{document}
\maketitle

\hrule

\begin{abstract}
Dimension reduction is an important tool for analyzing high-dimensional data. The predictor envelope is a method of dimension reduction for regression that assumes certain linear combinations of the predictors are immaterial to the regression. The method can result in substantial gains in estimation efficiency and prediction accuracy over traditional maximum likelihood and least squares estimates. While predictor envelopes have been developed and studied for independent data, no work has been done adapting predictor envelopes to spatial data. In this work, the predictor envelope is adapted to a popular spatial model to form the spatial predictor envelope (SPE). Maximum likelihood estimates for the SPE are derived, along with asymptotic distributions for the estimates given certain assumptions, showing the SPE estimates to be asymptotically more efficient than estimates of the original spatial model. The effectiveness of the proposed model is illustrated through simulation studies and the analysis of a geochemical data set.
\end{abstract}

\textit{Keyword:} Dimension Reduction,  Envelope model, Major chemical elements, Rare earth elements, Spatial correlation.

\vspace{0.1in}
\hrule

\doublespacing

\section{Introduction}
Spatial linear regression is a popular approach to estimate and interpret the relationship between different spatial variables \citep{ag1, ag2}, or to assist the interpolation/prediction of a spatial response, as in the regression-kriging procedure \citetext{\citealp{regressionkriging}; \citealp[Section~1.5]{stein}}. Estimates of the regression coefficients are often obtained via generalized least squares (GLS), which yields the maximum likelihood estimates for many spatial models \citep{mardiamarshall}. A large number of correlated predictors can result in high variance of the GLS estimates and subsequent poor prediction of the response. This is often addressed through dimension reduction, where the response is regressed on the predictors projected onto a lower-dimensional space. Many proposed methods of predictor dimension reduction for spatial regression are purely algorithmic. For example, \cite{spatialPCR} used principal component regression, while \cite{SPLS} and \cite{raca} used partial least squares. These methods use algorithms for dimension reduction designed for independent data, only addressing the spatial correlation on the back-end, when regressing the response on the reduced predictors. Others have used a Bayesian approach: \cite{SFnngp} used a spatial factor model, where the response is linked to the predictors through a small number of latent spatial variables. 
In this paper, we propose a likelihood-based approach, called the spatial predictor envelope.\par
The predictor envelope \citep{cookpredictor} is a likelihood-based method of dimension reduction for regression that assumes certain linear combinations of the predictors are immaterial to the regression and therefore can be excluded. This can result in massive gains in efficiency for parameter estimation and prediction accuracy. In addition to predictor envelopes, many other envelope models have been developed. \cite{lee2020review} provide an extensive review of advances in envelope methods. Predictor envelopes have only been developed and studied assuming independent observations, an assumption often untenable for spatial data \citep{cressie}. \cite{moradi} proposed a spatial response envelope for regression with a multivariate spatial response that provides substantial advantages over the original response envelope for spatial data, as well as over other contemporary spatial models.\par 
In this paper, we propose the spatial predictor envelope (SPE), providing a likelihood-based method for estimating the reduced predictor space and remaining model parameters. The model is an adaptation of the \lq separable\rq\ multivariate spatial model \citep{mardiagoodall}, where the covariance between the response and predictors is of reduced rank via the predictor envelope. Maximum likelihood estimates (MLEs) are derived and the model and corresponding estimates are demonstrated through simulation and a data analysis. The simulations and data analysis are carried out with a univariate response, performing envelope dimension reduction on a multivariate predictor. However, the theory remains the same for a multivariate response, with meaningful changes only in notation. Therefore, the theoretical developments of this paper allow for a multivariate response to keep the notation as general as possible.\par
Section \ref{review} gives a brief review of predictor envelopes and spatial regression. Section \ref{SPEmodel} presents the SPE model and parameter MLEs. Section \ref{theo} discusses the potential gains over the original spatial regression model and asymptotic distributions of the parameter estimates. Section \ref{sims} provides simulation studies and Section \ref{part3:data} demonstrates a real data analysis on a geochemical data set. Some technical details and proofs are given in the Appendix.


\section{Brief Review}\label{review}
In this section, the predictor envelope and spatial regression are briefly reviewed.
\subsection{Predictor Envelopes}
Consider the linear regression model 
\begin{equation}
\vY|\vX = \vmu_{Y|X} + \bbeta^T\vX + \vepsilon_{Y|X},
\end{equation}
where $\vY$ is an $r \times 1$ response, $\vX$ is a $p\times 1$ predictor, $\vmu_{Y|X}$ is an $r\times 1$ intercept vector,  $\bbeta$ is an $p\times r$ matrix of regression coefficients and $\vepsilon_{Y|X}$ is an $r\times 1$ error term with zero mean and constant variance. Under normality and iid assumptions, the MLE for $\bbeta$ is the ordinary least squares estimate. Predictor envelopes assume certain linear combinations of the predictors are invariant to changes in the response \citep{cookpredictor}. Let $\vY$ and $\vX$ be jointly multivariate normal. The envelope, $\env \subset \R^p$, is the smallest subspace of the predictor space such that
\begin{align}\label{EnvCond1}
\bQ_\env \vX|\vY = \vy_1 &\sim \bQ_\env \vX|\vY = \vy_2\\ 
\bP_\env \vX &\perp \bQ_\env \vX ,
\end{align}
where $\bP_\env$ and $\bQ_\env$ are projections onto the envelope and orthogonal complement $\env^\perp$ respectively. Orthogonality is with respect to the dot product, i.e. $\vec{A} \in \env, \vec{B} \in \env^{\perp}, \implies \vec{A}^T\vec{B} = 0$. The first condition declares the distribution of $\bQ_\env \vX$ to be invariant to changes in $\vY$. The second condition prevents $\bQ_\env \vX$ being affected by $\vY$ through correlation with $\bP_\env \vX$. Assume the envelope $\env$ is of dimension $u\leq p$. The previous conditions are equivalent to the parameterization:
\begin{align}
\bbeta &= \bGamma_1\boldeta \label{EnvPar1}\\
\var[\vX] = \bSigma_X &= \bGamma_1\bOmega_1\bGamma_1^T + \bGamma_0\bOmega_0\bGamma_0^T, \label{EnvPar2}
\end{align}
where $\bGamma_1$ is a $p\times u$ orthogonal basis for $\env$, $\bGamma_0$ is a $p\times (p-u)$ orthogonal basis for the orthogonal complement, $\boldeta$ is a $u\times r$ matrix of coordinates for $\bbeta$ with respect to $\bGamma_1$, and $\bOmega_1$ and $\bOmega_0$ are covariance matrices representing the variation within the envelope and orthogonal complement respectively. If $u=p$, no dimension reduction is performed and the model reverts to ordinary regression. If $u=0$, then $\bbeta = \bzero$ and no correlation exists between $\vY$ and $\vX$. Although particular bases $\bGamma_1$ and $\bGamma_0$ are not identifiable, the spaces they span are. Parameters $\bOmega_1$, $\bOmega_0$ and $\boldeta$ are identifiable given a particular choice of basis for the envelope and its orthogonal complement. For fixed $u$, the model parameters are estimated from the data via maximum likelihood. The estimated dimension $\hat{u}$ is selected via AIC, BIC or some other model selection technique. \par
Predictor envelopes improve estimation of $\bbeta$ over ordinary least squares by reducing the parameter count and excluding small eigenvalues of $\bSigma_X$ associated with dimensions within the orthogonal complement of $\env$ that are therefore immaterial to the regression and needlessly inflating the variance of estimates. Predictor envelopes deliver the greatest gains when $\|\bOmega_0\| \ll \|\bOmega_1\|$, where $\|\cdot\|$ is the spectral norm. More information on predictor envelopes can be found in \cite{cookpredictor}, along with comparisons to other predictor reduction techniques. \par
\cite{cookpredictor} derived MLEs for the predictor envelope parameters assuming independent observations. For spatial data, the assumption of independent observations is often untenable, as observations may be correlated based on their positions in space. \cite{moradi} developed a spatial response envelope model, showing improvement over the traditional response envelope model of \cite{cookresponse} for spatially correlated data. 


\subsection{Spatial Regression}\label{spatialregression}
Let $\vY(s)$ be an $r\times 1$ response and $\vX(s)$ a $p\times 1$ predictor, indexed by location $s\in D$, where $D$ is a spatial domain, typically a subset of $\R^2$. Assume the linear regression
\begin{equation}\label{fullmodelcond}
\vY(s)|\vX(s) = \vmu_{Y|X} + \bbeta^T\vX(s) + \vepsilon_{Y|X}(s),~~s\in D.
\end{equation}
Let the error term $\vepsilon_{Y|X}(s)$ be a Gaussian process with mean zero and separable covariance \citep{mardiagoodall}, i.e.
\begin{equation}
\cov[\vepsilon_{Y|X}(s), \vepsilon_{Y|X}(s')] = \rho_\theta(s,s')\bSigma_{Y|X},~~s,s'\in D,
\end{equation}
where $\bSigma_{Y|X}$ is an $r\times r$ covariance matrix and $\rho_\theta(s,s')$ is a valid univariate correlation function \citep{kernels}, dependent on finite set of parameters $\theta$. Given $n$ observations of the response and predictors at locations $s_1,\ldots,s_n\in D$, let
\begin{equation}\label{obsmatrices}
\bY = \begin{bmatrix} \vY(s_1)^T \\ \vdots \\ \vY(s_n)^T \end{bmatrix}, ~~~ \bX = \begin{bmatrix} \vX(s_1)^T \\ \vdots \\ \vX(s_n)^T\end{bmatrix}, ~~~ \tilde{\bX} = [\ones_n~\bX] 
\end{equation}
be $n\times r$, $n\times p$ and $n\times (p+1)$ matrices respectively, and $\brho(\theta)$ be an $n\times n$ matrix where $[\brho(\theta)]_{ij} = \rho_\theta(s_i, s_j)$. The MLEs for the intercept $\vmu_{Y|X}$ and regression coefficients $\bbeta$ are the generalized least squares (GLS) coefficients
\begin{equation}\label{GLS}
[\hat{\mu}_{Y|X}\ \hat{\bbeta}_\text{GLS}] = (\tX^T\brho(\hat{\theta})^{-1}\tX)^{-1}\tX^T\brho(\hat{\theta})^{-1}\bY,
\end{equation}
where $\hat{\theta}$ is the maximum likelihood estimate for $\theta$. The variance of $\hat{\bbeta}_\text{GLS}$ is inversely proportionate to the magnitude of the eigenvalues of $(\tX^T\brho(\hat{\theta})^{-1}\tX)$ (see Theorem \ref{theorem:potgains} in Section \ref{gainzz}). Small eigenvalues can result in poor estimation and inaccurate prediction of the response. Projecting the predictors onto a subspace not associated with some of these small eigenvalues while preserving information relevant to the regression results in improved estimation of the regression parameters and prediction of the response. \par
Predictor envelopes estimate a reducing subspace for the predictors through the likelihood, requiring joint modeling of the predictors. Let the joint process of $\vY(s), \vX(s)$ be a multivariate Gaussian process
\begin{equation} \label{fullmodel0}
\vZ(s) = \begin{bmatrix} \vY(s) \\ \vX(s) \end{bmatrix} = \vmu_Z+ \vepsilon_Z(s),~~s\in D,
\end{equation}
where $\vmu_Z = [\vmu_X^T~\vmu_Y^T]^T$ and $\vepsilon_Z(s)$ is a mean-zero Gaussian process with covariance
\begin{equation} \label{fulllcond}
\cov[\vepsilon_Z(s),\vepsilon_Z(s')] = \rho_\theta(s,s')\bSigma_Z, ~~~ \bSigma_Z = \begin{bmatrix} \bSigma_Y & \bSigma_{XY}^T \\ \bSigma_{XY} & \bSigma_X \end{bmatrix},
\end{equation}
where $\bSigma_{XY} = \bSigma_X\bbeta$. Note that this assumes $\vX(s)$ has the same spatial correlation function as $\vY(s)$. The MLE for $\bbeta$ is still the GLS coefficients of Equation \ref{GLS}, however the MLE for $\theta$ will be effected by the joint modeling of $\vX(s)$.

\section{Spatial Predictor Envelope}\label{SPEmodel}
In this section, we present the SPE model and parameter estimates.
\subsection{Model}
While the assumption of independent observations is convenient, it is problematic for data correlated over space \citep{cressie}. The predictor envelope can be adapted to spatial data by assuming the data follows the joint spatial model in Equation \ref{fullmodel0}, but with the envelope parameterization of Conditions \ref{EnvPar1} and \ref{EnvPar2}:
\begin{equation} \label{sepcond}
\begin{aligned}
\bSigma_X = \bGamma_1\bOmega_1\bGamma_1^T + \bGamma_0\bOmega_0\bGamma_0^T ~~ &\text{and} ~~ \bSigma_{XY} = \bGamma_1\bOmega_1\boldeta\\
\implies \bbeta &= \bGamma_1\boldeta,
\end{aligned}
\end{equation}
where the the envelope parameters are the same as those defined in Section \ref{review}. Throughout the paper, the model presented in Equation \ref{fullmodel0} that satisfies Conditions \ref{sepcond} is called the {\bf spatial predictor envelope} (SPE). If $u=p$, no dimension reduction is performed and the model is equivalent to the unconstrained, full spatial regression of Equation \ref{fullmodel0}. If $u=0$, then $\bSigma_{XY} = \bbeta = \bzero$, implying $\vY(s)$ and $\vX(s)$ are independent Gaussian processes. But if $0<u<p$, then $\bSigma_{XY}$ and $\bbeta$ are of reduced rank, with column vectors contained within the envelope space $\env = \text{span}(\bGamma_1)$. In this case, efficiency gains in estimation of $\bbeta$ are possible through a reduced parameter count and omission of small eigenvalues of $\bSigma_X$ (see Section \ref{gainzz}). If spatial parameters $\theta$ and envelope dimension $u$ are known, then the SPE is always asymptotically efficient versus the full spatial regression (see Section \ref{asy}). \par
The conditional distribution $\vY(s)|\vX(s)$ provides a more typical spatial regression format for the SPE
\begin{equation}
\begin{aligned}
\vY(s)|\vX(s) &= \vmu_{Y|X} + \bbeta^T\vX(s) + \vepsilon_{Y|X}(s) \\
&= \vmu_{Y|X} + (\bGamma_1\boldeta)^T\vX(s) + \vepsilon_{Y|X}(s),
\end{aligned}
\end{equation}
where $\vepsilon_{Y|X}(s)$ is a mean-zero Gaussian process of dimension $r$ with covariance
\begin{equation}
\begin{aligned}
\cov[\vepsilon_{Y|X}(s),\vepsilon_{Y|X}(s')] &= \rho_\theta(s,s')\bSigma_{Y|X} \\
\bSigma_{Y|X} &= \bSigma_Y - \bSigma_{XY}^T\bSigma_X^{-1}\bSigma_{XY}\\
&= \bSigma_Y - \boldeta^T\bOmega_1\boldeta.
\end{aligned}
\end{equation}
While this conditional formulation is useful for interpretation and prediction, the joint formulation of Equation \ref{fullmodel0} is used for parameter estimation, as the envelope parameters affect the distribution of $\vX(s)$.


\subsection{Maximum Likelihood Estimation} \label{mle}
Let $\bY$, $\bX$, $\tX$, and $\brho(\theta)$ be the observation matrices defined in Equation \ref{obsmatrices}, and let $\tY = [\ones\ \bY]$. For fixed envelope dimension $u$, the maximum likelihood estimates for the SPE are
\begin{equation}\label{SPEobj}
\begin{aligned}
\hat{\bGamma}_1,\hat{\theta} &= \argmin_{\bG_1,\phi} \{n\log|\bG_1^T\bS_{X|Y}(\phi)\bG_1| + n\log|\bG_1^T\bS_{X}^{-1}(\phi)\bG_1|\\ &+ n\log|\bS_{X}(\phi)| + n\log|\bS_{Y}(\phi)| + (r+p)\log|\brho(\phi)|\},
\end{aligned}
\end{equation}
where $\bG_1$ is a $p\times u$ orthogonal matrix representing $\bGamma_1$ during the optimization, $\phi$ is a potential set of spatial correlation parameters representing $\theta$ during the optimization, and 
\begin{equation}\label{samplecov}
\begin{aligned}
\bS_{X|Y}(\phi) &= \left(\bX - \tY(\tY^T\brho(\phi)^{-1}\tY)^{-1}\tY^T\brho(\phi)^{-1}\bX\right)^T\brho(\phi)^{-1} \left(\bX - \tY(\tY^T\brho(\phi)^{-1}\tY)^{-1}\tY^T\brho(\phi)^{-1}\bX\right)\\
\bS_{X}(\phi) &= \left(\bX - \ones(\ones^T\brho(\phi)^{-1}\ones)^{-1}\ones^T\brho(\phi)^{-1}\bX\right)^T\brho(\phi)^{-1} \left(\bX - \ones(\ones^T\brho(\phi)^{-1}\ones)^{-1}\ones^T\brho(\phi)^{-1}\bX\right)\\
\bS_{Y}(\phi) &= \left(\bY - \ones(\ones^T\brho(\phi)^{-1}\ones)^{-1}\ones^T\brho(\phi)^{-1}\bY\right)^T\brho(\phi)^{-1} \left(\bY - \ones(\ones^T\brho(\phi)^{-1}\ones)^{-1}\ones^T\brho(\phi)^{-1}\bY\right),
\end{aligned}
\end{equation}
$\bS_{X|Y}(\phi)$, $\bS_X(\phi)$ and $\bS_Y(\phi)$ being sample covariances adjusted for parameters $\phi$. The remaining parameter estimates are then
\begin{equation}\label{otherparams}
\begin{array}{c c c}
[\hat{\mu}_{Y|X}\ \hat{\bbeta}_\text{GLS*}] =(\tX^T\brho(\hat{\theta})^{-1}\tX)^{-1}\tX^T\brho(\hat{\theta})^{-1}\bY, & \hat{\boldeta} =  \hat{\bGamma}_1^T\hat{\bbeta}_\text{GLS*}, & \hat{\bbeta} = \hat{\bGamma}_1\hat{\boldeta}, \\
\hat{\bOmega}_1 = \hat{\bGamma}_1^T\bS_{X}(\hat{\theta})\hat{\bGamma}_1, & \hat{\bOmega}_0 = \hat{\bGamma}_0^T\bS_{X}(\hat{\theta})\hat{\bGamma}_0. & 
\end{array}
\end{equation}
An asterisk is included in $\hat{\bbeta}_\text{GLS*}$ to distinguish it from the traditional MLE, $\hat{\bbeta}_\text{GLS}$, because the estimate for $\theta$ is affected by joint estimation of $\bGamma_1$. Derivation of the MLEs can be found in Appendix \ref{mlea}. Estimated dimension $\hat{u}$ can be chosen via AIC, BIC or cross-validation. It is suggested that BIC or cross-validation be used when prediction of the response is the goal, and AIC when a more conservative dimension reduction is desired.\par

Objective Function \ref{SPEobj} requires optimization over $p\times u$ orthogonal $\bG_1$. The reparameterization of \cite{fastenv} can be applied for an unconstrained optimization: Without loss of generality, let the rows of $\bG_1$ be ordered such that the $u\times u$ matrix $\bG_{11}$ formed by the first $u$ rows is non-singular. Such an ordering must exist, because $\text{rank}(\bG_1) = u$. Then
\begin{equation}\label{repar}
\bG_1 = \begin{bmatrix} \bG_{11} \\ \bG_{12} \end{bmatrix} = \begin{bmatrix} \bI_{u} \\ \bA \end{bmatrix}\bG_{11} = \bC_{\bA}\bG_{11},
\end{equation}
where $\bA = \bG_{12}\bG_{11}^{-1}$ is an unconstrained $(p-u)\times u$ matrix and $\bC_{\bA} = [\bI_u\ \bA^T]^T$. Note that since the columns of $\bG_1$ are linear combinations of the columns of $\bC_A$, and vice-versa, the two matrices have the same column space. Making the substitution given by Equation \ref{repar} in Objective Function \ref{SPEobj} yields
\begin{equation}\label{SPEobj2}
\begin{aligned}
\hat{\bA},\hat{\theta} &= \argmin_{\bA,\phi} \{n\log|\bC_{\bA}^T\bS_{X|Y}(\phi)\bC_{\bA}| + n\log|\bC_{\bA}^T\bS_{X}^{-1}(\phi)\bC_{\bA}| - 2n\log|\bC_{\bA}^T\bC_{\bA}| \\ &+ n\log|\bS_{X}(\phi)| + n\log|\bS_{Y}(\phi)| + (r+p)\log|\brho(\phi)|\}.
\end{aligned}
\end{equation}
Then any off-the-shelf optimization routine can be applied over the unconstrained $\{\phi, \bA\}$, applying any necessary transformation to ensure $\phi$ is unconstrained. Typically the spatial correlation parameters are strictly positive, so a log-transformation will suffice. To reclaim the orthogonal parameterization $\hat{\bGamma}_1$ after the optimization is complete, take any orthogonal basis for the column space of $\bC_{\hat{\bA}}$, for example, the eigenvectors with non-zero eigenvalues of $\bC_{\hat{\bA}}\bC_{\hat{\bA}}^T$. The original, constrained Objective Function \ref{SPEobj} can also be optimized using routines with orthogonality constraints, such as \cite{wenyen} which is implemented in \lq R\rq\ package \lq rstiefel\rq\ \citep{rstiefel, Rcore}. 

\section{Theoretical Results}\label{theo}
In this section, we present the potential gains and asymptotic distributions of the SPE model parameters.
\subsection{Potential Gains}\label{gainzz}
General finite sample variances of $\hat{\bbeta}$ are not readily available for the SPE model. However, if we assume some of the nuisance parameters are known, finite sample variances can be derived that are instructive to the potential gains of the SPE model over the full spatial regression. 
\begin{theorem}\label{theorem:SPEvar}
Assume the envelope and corresponding basis $\bGamma_1$ are known, as well as spatial parameters $\theta$. The variance of the SPE estimates $\hat{\bbeta}$ are
\begin{equation}\label{envvar}
\var[\ve(\hat{\bbeta})] = \left(\frac{1}{n-u-2}\right)\bSigma_{Y|X} \otimes  \bGamma_1\bOmega_1^{-1}\bGamma_1^T.
\end{equation}
\end{theorem}
Proof of Theorem \ref{theorem:SPEvar} can be found in Appendix \ref{pga}. The following corollary provides the potential gains over the full spatial regression. 
\begin{corollary}\label{theorem:potgains}
Given the assumptions of Theorem \ref{theorem:SPEvar}, the variance of the full spatial regression (GLS) estimates $\hat{\bbeta}_\text{GLS}$ are
\begin{equation}\label{glsvar}
\var[\ve(\hat{\bbeta}_\text{GLS})] = \left(\frac{1}{n-p-2}\right) \bSigma_{Y|X} \otimes \bSigma_X^{-1},
\end{equation}
which implies the variance of the SPE estimates can be written as
\begin{equation}\label{varcompare}
\var[\ve(\hat{\bbeta})] = \left(\frac{n-p-2}{n-u-2}\right)\var[\ve(\hat{\bbeta}_\text{GLS})] -  \left(\frac{1}{n-u-2}\right)\bSigma_{Y|X} \otimes \bGamma_0\bOmega_0^{-1}\bGamma_0^T.
\end{equation}
\end{corollary}
By the previous corollary, the potential gains of the SPE over GLS are proportionate to the parameter reduction and inversely proportionate to the magnitude of $\bOmega_0$. While gains made through parameter reduction can be significant, especially for relatively small sample sizes, the greatest gains are possible via the omission of small eigenvalues in $\bOmega_0$.

\subsection{Asymptotic Distributions}\label{asy}
In this section, it is assumed that spatial parameters $\theta$ are known, and the asymptotic distributions of the SPE parameter estimates are studied and compared to that of the unconstrained, full spatial regression model of Equation \ref{fullmodel0} with no dimension reduction. Under the assumption of known $\theta$, the asymptotic variances of the SPE are equivalent in form to those of the traditional predictor envelope. \par
Asymptotic variance of the estimates are not directly available through inversion of the Fisher Information matrix because $\bGamma_1$ is over parameterized: there are multiple possibilities for the $p\cdot u$ entries in matrix $\bGamma_1$ that lead to an orthogonal basis for the same space, and therefore give equivalent models. The techniques of \cite{cookresponse,cookpredictor} are followed, applying the results of \cite{shapiro} for deriving asymptotic variances of over-parameterized models. Let 
\[\psi= \begin{bmatrix}  \vech(\bSigma_{Y|X}) \\ \ve(\bGamma_1) \\ \vech(\bOmega_1) \\ \vech(\bOmega_0) \end{bmatrix} \]  
and $h(\psi)$ be a vector of the estimable parameters
\[h(\psi)= \begin{bmatrix} \vech(\bSigma_{Y|X}) \\ \vech(\bSigma_X) \\ \ve(\bbeta) \end{bmatrix}. \]  
Intercept $\vmu_Z$ is omitted because its estimates are asymptotically independent of the others. Let $\bH$ be the gradient matrix $[\bH]_{ij} = \frac{\partial h_i}{\partial \psi_j}$ and let $\bJ_F$ be the Fisher Information matrix of $h(\psi)$ under the unconstrained model of Equation \ref{fullmodel0}. Under the assumption of known $\theta$, the parameter estimates for the unconstrained model are consistent and asymptotically normal with variance $\bJ_F^{-1}$ (more details in Appendix \ref{avara}). Then by \cite[Proposition~4.1]{shapiro} the following theorem results
\begin{theorem}\label{theorem:avar}
Assume spatial parameters $\theta$ are known. Then $h(\hat{\psi}) \rightarrow N(h(\psi), \bJ_\text{SPE}^{-1})$, where
\begin{equation}\label{avarProj}
\text{avar}[\sqrt{n}h(\psi)] \equiv \bJ_\text{SPE}^{-1} = \bH(\bH^T\bJ_F\bH)^\dagger\bH^T,
\end{equation}
where $\dagger$ is the Moore-Penrose inverse. In particular, the asymptotic variances for parameter estimates $\bbeta$ are
\begin{equation}\label{avarbeta}
\text{avar}[\sqrt{n}\ve(\hat{\bbeta})] = \bSigma_{Y|X}\otimes\bGamma_1\bOmega_1^{-1}\bGamma_1^T + (\boldeta^T\otimes\bGamma_0)\bM^{-1}(\boldeta\otimes\bGamma_0^T),
\end{equation}
where 
\[\bM = \boldeta\bSigma_{Y|X}^{-1}\boldeta^T \otimes \bOmega_0 + \bOmega_1\otimes\bOmega_0^{-1} + \bOmega_1^{-1}\otimes \bOmega_0 - 2\bI_u \otimes \bI_{p-u} .\]
\end{theorem}

These asymptotic variances are the same as those of the traditional predictor envelope \citep{cookpredictor}. Note that when these formulae are used in practice, the plug-in estimates of the asymptotic variance will use the MLEs for the SPE model, which can differ greatly than those of the traditional predictor envelope model. \par

The next theorem declares the SPE estimates are asymptotically efficient compared to the full spatial regression (GLS) estimates.
\begin{theorem}\label{theorem:efficient} 
Let $J_F^{-1}$ be the asymptotic variance for the full regression model and $J_\text{SPE}^{-1}$ be the asymptotic variance for the SPE estimates given by Equation \ref{avarProj}. Then
\[\bJ_F^{-1} -  \bJ_\text{SPE}^{-1} \geq 0. \]
\end{theorem} 
Proof of Theorem \ref{theorem:efficient} is in the Appendix. In Section \ref{sims}, the asymptotic distributions for $\hat{\bbeta}$ given by Equation \ref{avarbeta} are compared to the empirical distribution of simulation estimates of $\bbeta$, where $\theta$ is estimated from the data for every sample. The two distributions match each other well, suggesting the above asymptotic variances are useful in practice. \par

\section{Simulations}\label{sims}
Simulations 1 and 2 present scenarios advantageous to the SPE, where Simulation 1 assumes envelope dimension $u$ is known and fixed to study estimation of $\bGamma_1$ and Simulation 2 assumes $u$ is unknown to study dimension selection and estimation of $\bbeta$. Simulations 3 and 4 present scenarios where the SPE provides no advantage over traditional spatial regression (GLS). Simulation 5 compares the asymptotic distributions of Equation \ref{avarbeta} to the empirical distribution of the SPE regression estimates $\hat{\bbeta}$.\\
\textbf{Simulation 1:} The accuracy of envelope subspace estimation from objective function \ref{SPEobj} is compared to the envelope subspace estimation in \cite{cookpredictor, fastenv}. The response $\vY(s)$ and predictors $\vX(s)$ are simulated at $n$ locations $s_1,\ldots, s_n$ according to the SPE model, given by Equation \ref{fullmodel0} subject to the envelope conditions \ref{sepcond}. For this first simulation, envelope dimension $u$ is assumed to be known. The data dimensions are $p = 10$, $u = 3$, $r=1$, and the model parameters are $\bOmega_1 = \text{diag}\{\exp(-1^{2/3}),\exp(-2^{2/3}),\exp(-3^{2/3}\}$, $\bOmega_0 = \text{diag}\{\exp(-4^{2/3}, -5^{2/3}, \ldots, -10^{2/3}\}$, $\boldeta = [1\ 1\ 1]^T$, $\bSigma_{Y|X} = 0.05$. These particular data dimensions are chosen to mirror the data analysis in Section \ref{part3:data} and parameters $\bOmega_1,\bOmega_0$ are chosen such that $\|\bOmega_0\| \ll \|\bOmega_1\|$, presenting a scenario advantageous to predictor envelopes.  Basis $\bGamma_1$ is randomly generated every iteration according to the methods of \cite{randortho} using the \lq R\rq\ package \lq pracma\rq\ \citep{pracma, Rcore}, and spatial coordinates are randomly generated according to a uniform distribution on $[0,1]\times[0,1]$ every iteration. An exponential correlation function with white noise is used
\begin{equation}\label{expocorrelation}
\rho(s,s') = \tau 1_{s=s'} + (1 - \tau)\exp\left(-\frac{\|s-s'\|}{\lambda}\right), 
\end{equation}
where $\tau \in [0,1]$ controls the proportion of noise relative to the spatial variation and $\lambda \in \R^+$ is a range parameter. We used $\tau = 0.1$ and $\lambda = 0.3$, giving the data substantial spatial correlation. The simulation is repeated 500 times for $n=50,100,200$ respectively. Models are compared by the angle between the estimated and true envelope space, $\angle\{\text{span}(\hat{\bGamma}_1),\ \text{span}(\bGamma_1)\}$, and the squared Euclidean distance between the estimated and true regression coefficients $\|\hat{\bbeta} - \bbeta\|^2$ . The angle between two subspaces of equal dimension with orthogonal bases $\bA$ and $\bB$ respectively is computed as the arc-cosine of the largest singular value of $\bA^T\bB$, which is equivalent to the largest principal angle between the subspaces \citep{subspaceangles}. Simulation results are provided in Table \ref{sim1}. The SPE improves estimation of $\text{span}(\bGamma_1)$ and $\bbeta$ over the traditional predictor envelope by accounting for the spatial correlation. The SPE improves estimation of $\hat{\bbeta}$ over the full spatial regression by excluding the small eigenvalues in $\bOmega_0$. 

\begin{table}[!htbp]
\caption{Average error between estimated and true subspace, $\angle\{\text{span}(\hat{\bGamma}_1),\ \text{span}(\bGamma_1)\}$, and average squared error of regression coefficients, $\|\hat{\bbeta} - \bbeta\|^2$. Standard deviations of the average are in parentheses.}  
\centering
\begin{tabular}{ccccc} 
\toprule
$n$ & Error Metric & SPE & PE & GLS \\ \midrule
50   &$\angle\{\text{span}(\hat{\bGamma}_1),\ \text{span}(\bGamma_1)\}$ & 0.082 (0.001) & 0.130 (0.003) & NA \\ 
       &$\|\hat{\bbeta} - \bbeta\|^2$ & 0.144 (0.009) & 0.361 (0.018) & 0.530 (0.017) \\ 
100  &$\angle\{\text{span}(\hat{\bGamma}_1),\ \text{span}(\bGamma_1)\}$ & 0.054 ($<$ 0.001) & 0.114 (0.002) & NA \\ 
       &$\|\hat{\bbeta} - \bbeta\|^2$ & 0.047 (0.002) & 0.237 (0.012) & 0.206 (0.006) \\
200    &$\angle\{\text{span}(\hat{\bGamma}_1),\ \text{span}(\bGamma_1)\}$ & 0.038 ($<$ 0.001) & 0.108 (0.002) & NA \\ 
         &$\|\hat{\bbeta} - \bbeta\|^2$ & 0.022 ($<$ 0.001) & 0.175 (0.007) & 0.093 (0.003) \\ \bottomrule
\end{tabular}

\label{sim1}
\end{table} 

\noindent\textbf{Simulation 2:} Data is simulated according to the same parameters as Simulation 1, but $u$ is estimated via BIC. Table \ref{sim2} gives average errors $\|\hat{\bbeta} -\bbeta\|^2$ and Figure \ref{fig:sim2} in Appendix \ref{sra} gives histograms of selected envelope dimension $\hat{u}$. The SPE improves estimation of $\hat{\bbeta}$ over the traditional predictor envelope and the full spatial regression, and improves selection of $\hat{u}$ over the traditional predictor envelope. It appears that as $n \rightarrow \infty$, the SPE selects to true dimension via BIC with probability approaching one. 

\begin{table}[!htbp]
\caption{Average error $\|\hat{\bbeta} - \bbeta\|^2$ with standard deviations in parentheses. Simulation scenario where $u < p$ and $\|\bOmega_0\| \ll \|\bOmega_1\|$.} 
\centering
\begin{tabular}{cccc}
\toprule
$n$ & SPE & PE & GLS \\ \midrule
50 & 0.221 (0.015) & 0.625 (0.030) & 0.530 (0.017) \\ 
100 & 0.075 (0.004) & 0.443 (0.020) & 0.206 (0.006) \\ 
200 & 0.029 (0.002) & 0.370 (0.014) & 0.093 (0.002)\\ \bottomrule
\end{tabular}
\label{sim2}
\end{table}

\noindent\textbf{Simulation 3:} Data is simulated according to the same parameters as Simulations 1 and 2, but $u = p = 10$, $\bOmega_1 =  \text{diag}\{\exp(-1^{2/3}),\exp(-2^{2/3}),\ldots, \exp(-10^{2/3}\}$ and $\boldeta = [1 \cdots 1]^T$, meaning $\bSigma_X$ has small eigenvalues, but there is no proper envelope subspace. Envelope dimension is unknown and estimated via minimum BIC, giving the SPE an opportunity to falsely estimate a reducing subspace. Table \ref{sim3} gives average errors $\|\hat{\bbeta} -\bbeta\|^2$ and Figure \ref{fig:sim3} in Appendix \ref{sra} gives histograms of selected envelope dimension $\hat{u}$. The SPE and GLS estimates perform similarly. The SPE often falsely estimates a reducing subspace via minimum BIC, but apparently not to the detriment (on average) of the regression estimates $\hat{\bbeta}$.

\begin{table}[!htbp]
\caption{Average error $\|\hat{\bbeta} - \bbeta\|^2$ with standard deviations in parentheses. Simulation scenario where $u = p$.} 
\centering
\begin{tabular}{cccc}
\toprule
$n$ & SPE & PE & GLS \\ \midrule
50 & 0.517 (0.018) & 0.970 (0.031) & 0.530 (0.017) \\ 
100 & 0.213 (0.007) & 0.629 (0.021) & 0.206 (0.006) \\ 
200 & 0.091 (0.003) & 0.475 (0.015) & 0.093 (0.002)\\ \bottomrule
\end{tabular}
\label{sim3}
\end{table}

\noindent\textbf{Simulation 4:} Data is simulated according to the same parameters as Simulations 1 and 2, but $\bOmega_1 = \text{diag}\{\omega_1,\omega_2,\omega_3\}$ and $\bOmega_0 = \text{diag}\{\omega_4,\ldots,\omega_{10}\}$, where $\omega_j,\ j=1,\dots,10$ are randomly generated according to a uniform distribution of $[0, 1]$ every iteration. In this way, the magnitudes of $\|\bOmega_1\|$ and $\|\bOmega_0\|$ are on average comparable. Table \ref{sim4} gives average errors $\|\hat{\bbeta} -\bbeta\|^2$ and Figure \ref{fig:sim4} in Appendix \ref{sra} gives histograms of selected envelope dimension $\hat{u}$. The SPE provides no significant advantage over the GLS estimates in this scenario, due to the comparable magnitudes of $\bOmega_1\|$ and $\|\bOmega_1\|$.

\begin{table}[!htbp]
\caption{Average error $\|\hat{\bbeta} - \bbeta\|^2$ with standard deviations in parentheses. Simulation scenario where $u < p$ and $\|\bOmega_1\| \approx \|\bOmega_0\|$.} 
\centering
\begin{tabular}{cccc}
\toprule
$n$ & SPE & PE & GLS \\ \midrule
50 & 0.105 (0.015) & 0.189 (0.025) & 0.122 (0.016) \\ 
100 & 0.040 (0.004) & 0.115 (0.013) & 0.042 (0.003) \\ 
200 & 0.019 (0.003) & 0.151 (0.037) & 0.020 (0.002)\\ \bottomrule
\end{tabular}
\label{sim4}
\end{table}

\begin{figure}[!htbp]
	\centering
	\begin{subfigure}{0.3\textwidth}
		\centering
		\includegraphics[width = \textwidth,keepaspectratio]{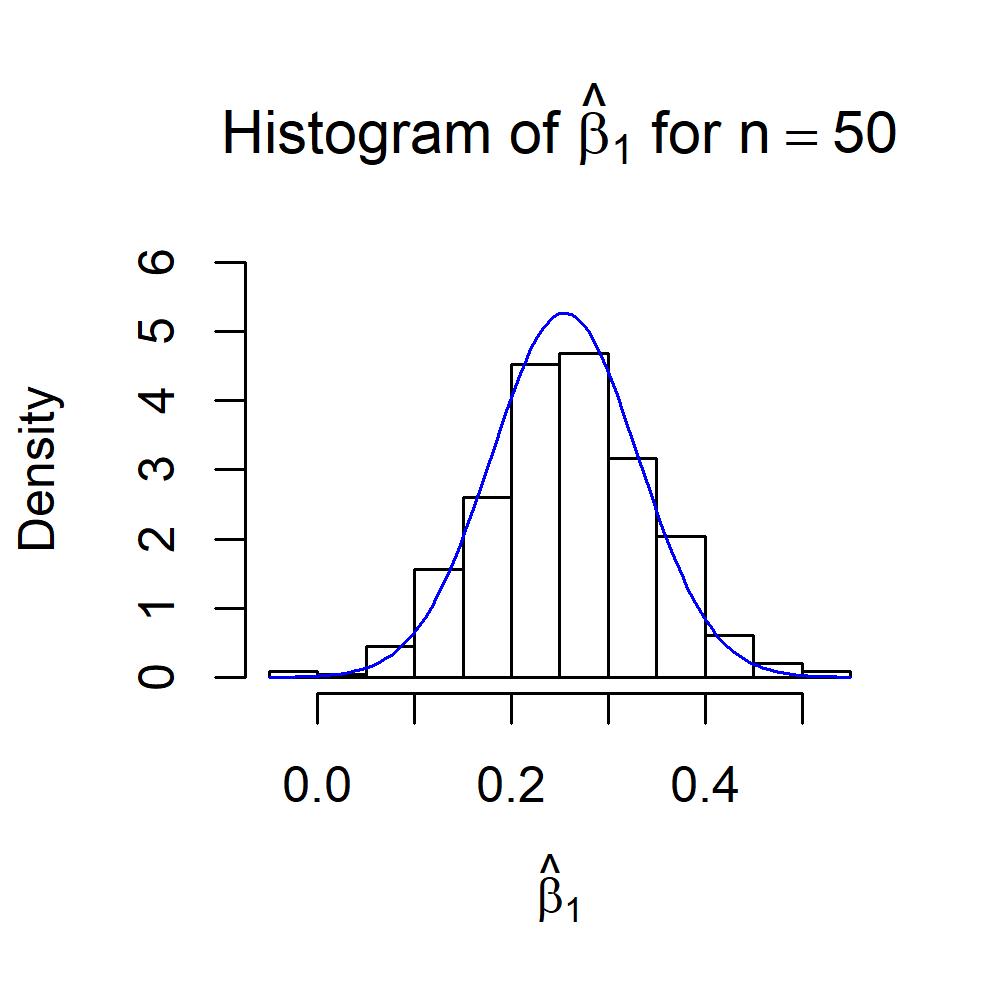}
	\end{subfigure}
	\begin{subfigure}{0.3\textwidth}
		\centering
		\includegraphics[width = \textwidth,keepaspectratio]{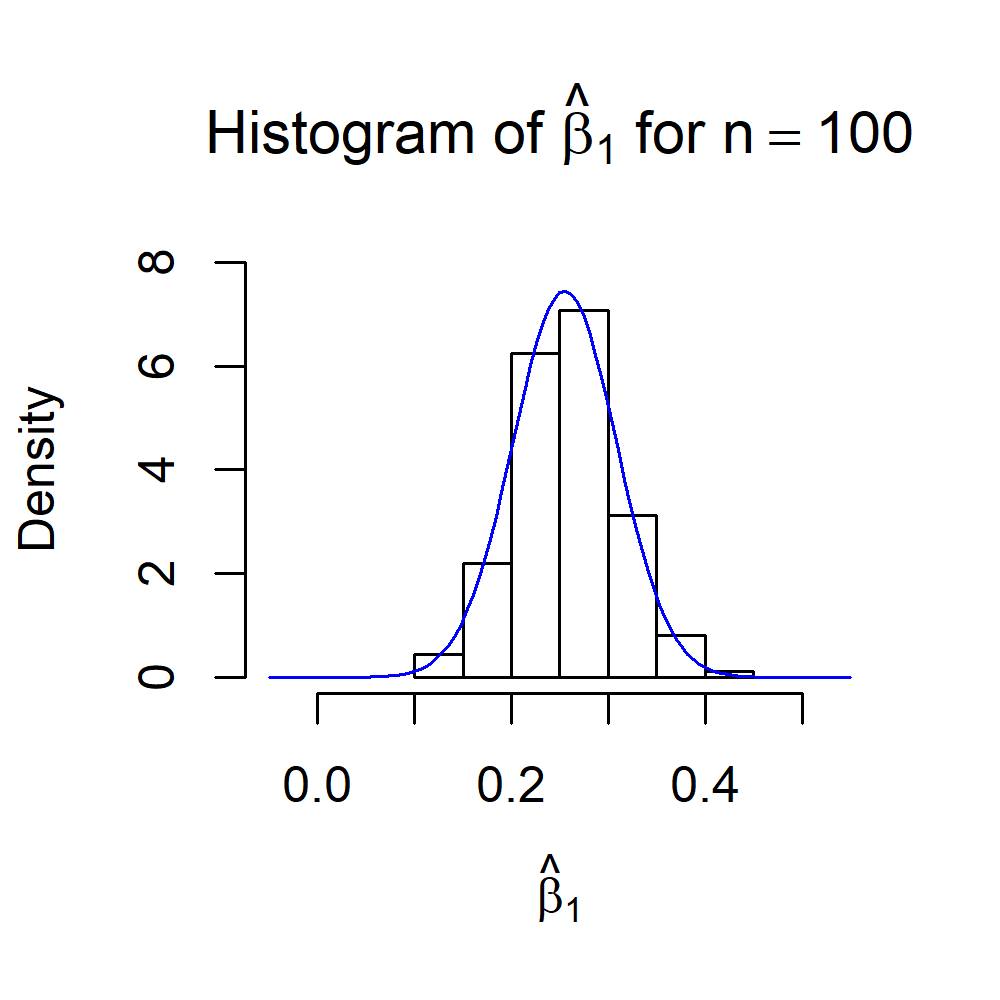}
	\end{subfigure} 
	\begin{subfigure}{0.3\textwidth}
		\centering
		\includegraphics[width = \textwidth,keepaspectratio]{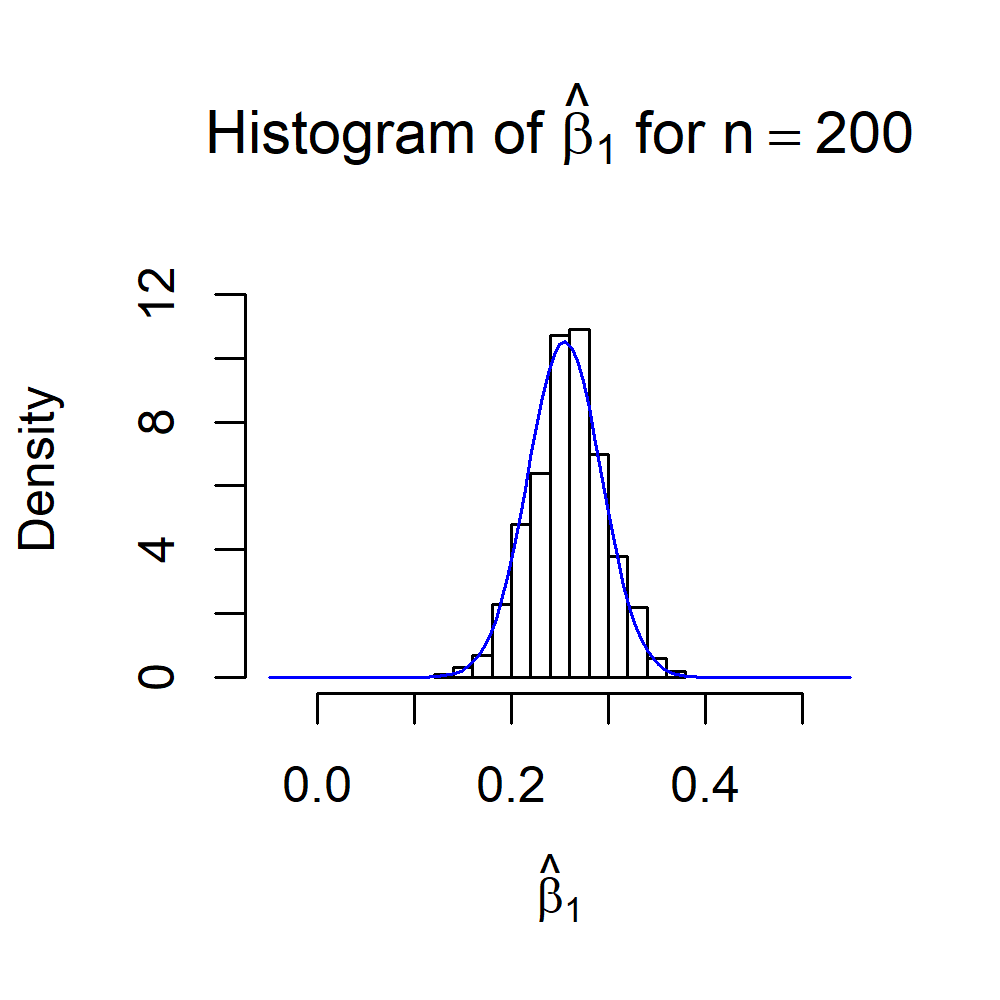}
	\end{subfigure}  
\caption{Histograms for the first estimated regression coefficient $\hat{\bbeta}_1$ across 500 simulation iterations with the asymptotic normal density given by Equation $\ref{avarbeta}$ plotted over-top.}
\label{avarsims}
\end{figure}

\noindent\textbf{Simulation 5:} The asymptotic distributions for $\hat{\bbeta}$ given by Equation \ref{avarbeta} are compared to the empirical distribution of simulation estimates of $\bbeta$ where $\theta$ is estimated from the data for every sample. The model parameters are the same as Simulations 1 and 2, except that the basis $\bGamma_1$ and the spatial coordinates are randomly generated once and held fixed throughout all 500 simulation iterations. In this way, the variance of $\hat{\bbeta}$ for a fixed model can be studied. For the same reason, $u$ is considered known and fixed. The simulation is repeated for $n=50,100,200$. Figure \ref{avarsims} gives histograms for the first estimated regression coefficient $\hat{\bbeta}_1$ with the asymptotic normal density given by Equation $\ref{avarbeta}$ plotted over-top. The empirical distributions and asymptotic distributions agree well, suggesting that the asymptotic distributions are useful in practice. 

\section{Data Analysis}\label{part3:data}

\begin{figure}[h]
\centering
\includegraphics[scale = 0.5]{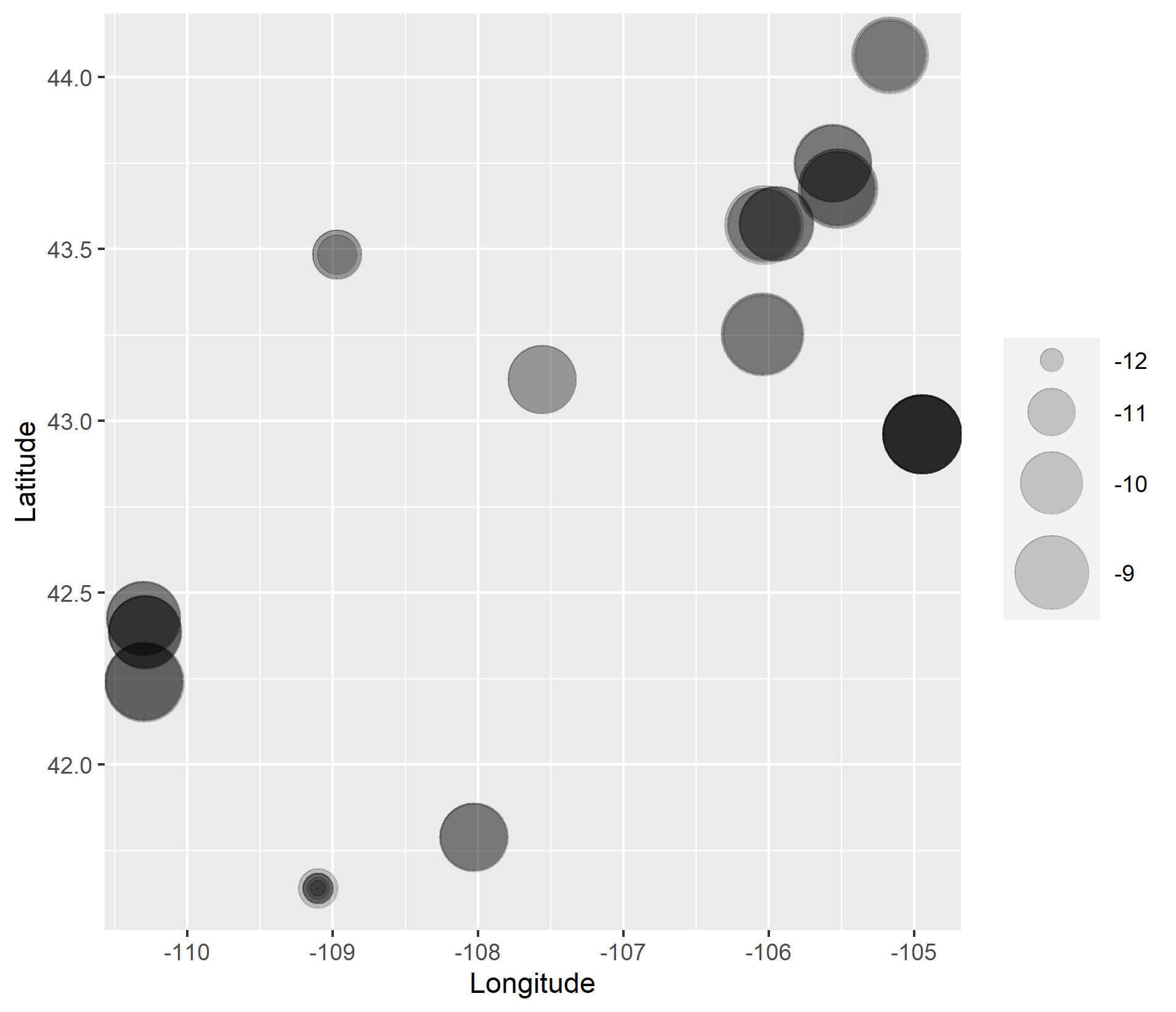}
\caption{Locations and levels of sampled total REE log-concentrations.}
\label{fig:REheatmap}
\end{figure}

We analyze a geochemical data set to illustrate the spatial predictor envelope. The data set contains 53 samples of the concentrations of 11 major chemical elements and 17 rare earth elements (REEs) at 14 different locations across an oil and gas reserve in Wyoming \citep{rareearthdata}. The data was collected to assess how rock type / major chemical element concentrations affected the concentration of REEs. REEs are critical to many modern technologies and industrial applications, but economically viable concentrations of REEs are uncommon and tend to occur in uncommon rock types \citep{usgs}. There is interest in identifying general mineral characteristics that are correlated with high REE concentrations and using these correlations to identify REE mining prospects; see for example \cite{utah}. We follow the techniques of \cite{aitchison} for compositional data, analyzing the log-ratios of the components. The log-concentrations of the 17 REEs are highly positively correlated, with over 95\% of the variation explained by the first principal component. With few exceptions, REEs do not occur individually but vary in unison \citep{REEmineralogy}. Therefore, we used the total log-concentration of REEs as the univariate regression response, as it simplifies the model and remains in line with the data analysis goals. We used the log-concentrations of the major chemical elements as a multivariate regression predictor and the SPE to estimate a reducing subspace and linear relationship between the total REE log-concentration and major chemical element log-concentrations. Appendix \ref{dataa} contains more details and justifications on the choice of variables and data transformations. Figure \ref{fig:REheatmap} gives the locations and levels of sampled total REE log-concentrations.\par
\begin{figure}[h]
\centering
\includegraphics[scale = 0.5]{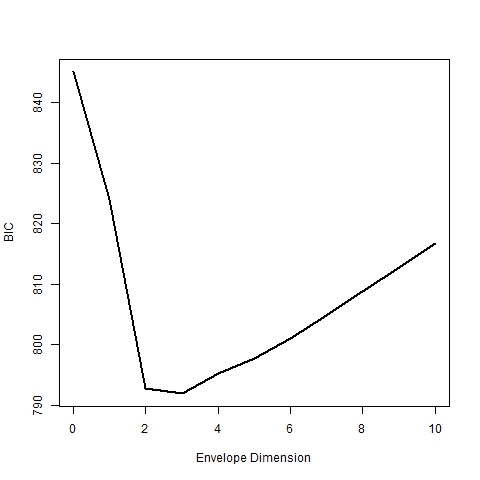}
\caption{BIC scores for the SPE for $u=0,\ldots,p$. The SPE achieves a minimum of $792$ at $u = 3$.}
\label{bic}
\end{figure}

An exponential correlation function with white noise (Equation \ref{expocorrelation}) was used for all correlation functions. The SPE was fit to the data for $u = 0,\ldots,p$ and the BIC scores were compared. Figure \ref{bic} gives a plot of BIC scores against dimension $u$.
The SPE achieves a minimum BIC of $792$ at $u=3$. The BIC for the full spatial regression model is $817$. For reference, the traditional predictor envelope of \cite{cookpredictor}, designed for independent observations, achieves a minimum BIC of $1197$, also at $u=3$. The substantially higher BIC score of the traditional predictor envelope when compared to the spatial models suggests accounting for the spatial correlation in this data is crucial. Table \ref{beta} gives the estimated coefficients, asymptotic standard errors and Z-scores $\hat{\bbeta}_i/\sqrt{\text{avar}[\hat{\bbeta}_i]}$ for the SPE model with $u=3$. Significant relationships are indicated between several major chemical concentrations and the total REE log-concentration, for example, a positive relationship for potassium oxide (K2O) and phosphorus pentoxide (P2O5) and a negative relationship for magnesium oxide (MgO) and silicon dioxide (SiO2). Alkaline igneous rocks are enriched in potassium and known to be hosts of REE deposits, providing an explanation for the strong positive relationship between potassium oxide and total REE log-concentration \citep{alkaline}.

\begin{table}[h]
\caption{Estimated coefficients, asymptotic standard errors and Z-scores $\hat{\bbeta}_i/\sqrt{\text{avar}[\hat{\bbeta}_i]}$ for each of the major chemical components.}
\centering
\begin{tabular}{ccccccccccc}
\toprule
Chemical & CaO & Fe2O3 & K2O & LOI & MgO & MnO & Na2O & P2O5 & SiO2 & TiO2 \\ \midrule
$\hat{\bbeta}_i$ & -0.10 & -0.12 & 0.36 & -0.05 & -0.21 & 0.03 & 0.02 & 0.22 & -0.29 & 0.10 \\ 
$\sqrt{\text{avar}[\hat{\bbeta}_i]}$ & 0.08 & 0.09 & 0.08 & 0.09 & 0.07 & 0.07 & 0.06 & 0.09 & 0.10 & 0.10 \\ 
Z-score & -1.28 & -1.35 & 4.50 & -0.58 & -3.13 & 0.38 & 0.45 & 2.57 & -2.73 & 1.03 \\ \bottomrule
\end{tabular}
\label{beta}
\end{table}

To compare the predictive performance of the SPE against traditional spatial regression (GLS), a cross-validation study was performed on the data. A hundred iterations of 10-fold cross-validation was performed, with predictions of the test data performed using the conditional expectation
\[\hat{Y}_\text{test} = E[Y_\text{test}|Y_\text{train},\hat{\bbeta},\hat{\theta}] = X_\text{test}\hat{\bbeta} + E[\epsilon_\text{test}|\epsilon_\text{train},\hat{\theta}]\] 
where the conditional expectation of the error terms is given by the kriging equation \citep{matheron, cressie}. The predictions were compared by their mean squared prediction error (MSPE), calculated as
\[\text{MSPE} = \frac{1}{N}\sum_{i=1}^{N}(\hat{Y}_{\text{test},i} - Y_{\text{test},i})^2, \]
where $N$ is the number of predicted observations. The average MSPE for the SPE and spatial regression was 0.09 and 0.16 respectively, showing significant improvement in prediction through the dimension reduction of the predictors. For context, the sample variance of the response, $\frac{1}{n}\sum_{i=1}^n(Y_i-\bar{Y})^2$, is 1.31, meaning both methods predicted a substantial portion of the variation in the response.

\section{Discussion}

The proposed spatial predictor envelope provides an effective tool for multivariate spatial data. By reducing the dimension of the predictors through a likelihood-based approach, the SPE can improve estimation of the regression coefficients and prediction of the response over traditional spatial regression. The SPE improves estimation/prediction over the original predictor envelope for spatial data by accounting for spatial correlation. \par
In this paper, the asymptotic distribution for the estimable parameters is given for known spatial correlation parameters $\theta$. The difficulty in deriving asymptotic distributions for unknown $\theta$ are two-fold. First, to apply the results of \cite{shapiro}, a discrepancy function satisfying certain regularity conditions must be devised such that the parameters that minimize the discrepancy function are equivalent to the parameters that maximize the likelihood. Second, the results of \cite{shapiro} hinge upon the parameter estimates of the full, unconstrained model being consistent and asymptotically normal. This is a difficult problem in spatial statistics, where the answer often depends on the sampling scheme of the locations $s_1,\ldots,s_n$ and the spatial correlation function \citep{mardiamarshall, zhangmatern}. \par
The assumption of normality may be too restrictive for some applications. \cite{cookfoundations} discuss envelope models in a generalized linear model context. Generalized linear mixed models are popular for modeling non-Gaussian spatial data, see for example \cite{glmm}. Expanding spatial envelopes to the general linear mixed model setting would be a worthy task.

\bibliographystyle{plainnat}
\bibliography{SPEbib_V3}

\section{Appendix}
\subsection{Maximum Likelihood Estimation}\label{mlea}

Let $\bY$ and $\bX$ be $n\times r$ and $n \times p$ matrices of observations at locations $s_1,\ldots,s_n$ and $\brho(\theta)$ be the $n\times n$ correlation matrix where $\brho(\theta)_{ij} = \rho_\theta(s_i,s_j)$. Let $\ones$ be a $n\times 1$ vector of ones and $\tY = [\ones\ \bY]$, $\tX = [\ones\ \bX]$. Assume the envelope dimension $u$ is known. \par
Derivation of the estimates is easiest if the joint log-likelihood $L(\bX,\bY)$ is decomposed into $L(\bX,\bY) = L(\bX|\bY) + L(\bY)$. First, see that
\begin{align*}
\bSigma_{X|Y} &= \bSigma_X - \bSigma_{XY}\bSigma_Y^{-1}\bSigma_{XY}^T\\
&= \bGamma_1\bOmega_1\bGamma_1^T + \bGamma_0\bOmega_0\bGamma_0^T - \bGamma_1\bOmega_1\boldeta\bSigma_Y^{-1}(\bGamma_1\bOmega_1\boldeta)^T\\
&= \bGamma_1(\bOmega_1 - (\bOmega_1\boldeta)\bSigma_Y^{-1}( \bOmega_1\boldeta)^T)\bGamma_1^T +  \bGamma_0\bOmega_0\bGamma_0^T \\
&\equiv  \bGamma_1\bOmega_1^*\bGamma_1^T +  \bGamma_0\bOmega_0\bGamma_0^T,
\end{align*}
showing that $\bSigma_{X|Y}$ has the same decomposition as $\bSigma_X$, but the material variation $\bOmega_1^*=\bOmega_1 - (\bOmega_1\boldeta)\bSigma_Y^{-1}( \bOmega_1\boldeta)^T$ has been reduced by the variation explained by $\bY$. Next, given the regression equation $\vX(s)|\vY(s) = \vmu_{X|Y} + \bbeta_X^T\vY + \vepsilon_{X|Y}(s)$ the reverse regression coefficients $\bbeta_X$ can be written as
\begin{align*}
\bbeta_X &= \bSigma_Y^{-1}\bSigma_{XY}^T\\
&= \bSigma_Y^{-1}(\bGamma_1\bOmega_1\boldeta)^T\\
&\equiv (\bGamma_1\boldeta_X)^T,
\end{align*}
where $\boldeta_X = \bOmega_1\boldeta\bSigma_Y^{-1}$, showing that $\bbeta_X$ has the same form as $\bbeta$. \par
Using properties of the Kronecker product, $|\bSigma_{X|Y} \otimes \brho(\theta)| = |\bSigma_{X|Y}|^n|\brho(\theta)|^p$. Finally, because $\bGamma_1,\ \bGamma_0$ are orthogonal, $|\bSigma_{X|Y}| = |\bOmega_1^*||\bOmega_0|$. \par
Now prepared to embark, the log-likelihood can be written as

\begin{equation*}
\begin{aligned}
L(\bX,\bY) =& -\frac{n}{2}|\bSigma_{X|Y}|  - \frac{1}{2}\ve(\bX - \ones\vmu_{X|Y}^T - \bY\bbeta_X)^T(\bSigma_{X|Y}\otimes\brho(\theta))^{-1}\ve(\bX - \ones\vmu_{X|Y}^T - \bY\bbeta_X) \\
&-\frac{n(p+r)}{2}\log(2\pi) - \frac{p+r}{2}|\brho(\theta)|\\ &-\frac{n}{2}|\bSigma_{Y}| -\frac{1}{2}\ve(\bY - \ones\vmu_{Y}^T)^T(\bSigma_{Y}\otimes\brho(\theta))^{-1}\ve(\bY - \ones\vmu_{Y}^T)\\[10pt]
=& -\frac{n}{2}|\bOmega_1^*| -\frac{1}{2}\ve(\bX\bGamma_1 - \ones\vmu_{X|Y}^T\bGamma_1 - \bY\boldeta_X^T)(\bOmega_1^*\otimes \brho(\theta))^{-1}\ve(\bX\bGamma_1 - \ones\vmu_{X|Y}^T\bGamma_1 - \bY\boldeta_X^T)\\ 
&-\frac{n}{2}|\bOmega_0|  -\frac{1}{2}\ve(\bX\bGamma_0 - \ones\vmu_{X|Y}^T\bGamma_0)(\bOmega_0\otimes \brho(\theta))^{-1}\ve(\bX\bGamma_0 - \ones\vmu_{X|Y}^T\bGamma_0) \\
&-\frac{n(p+r)}{2}\log(2\pi) - \frac{p+r}{2}|\brho(\theta)|\\ &-\frac{n}{2}|\bSigma_{Y}| -\frac{1}{2}\ve(\bY - \ones\vmu_{Y}^T)^T(\bSigma_{Y}\otimes\brho(\theta))^{-1}\ve(\bY - \ones\vmu_{Y}^T).
\end{aligned}
\end{equation*}
Let $\vmu_1 = \bGamma_1\vmu_{X|Y}$ and $\vmu_0 = \bGamma_0\vmu_{X|Y}$. Taking derivatives and equating to zero shows maximums are achieved at
\begin{align*}
 \hat{\bB}_1 \equiv \begin{bmatrix} \hat{\mu}_1 \\ \hat{\boldeta}_X \end{bmatrix} &= \left((\tY^T\brho(\theta)^{-1}\tY)^{-1}\tY^T\brho(\theta)^{-1}\bX\bGamma_1\right)^T\\
\hat{\mu}_0 &= \left((\ones^T\brho(\theta)^{-1}\ones)^{-1}\ones^T\brho(\theta)^{-1}\bX\bGamma_1\right)^T\\
\hat{\mu}_Y &= \left((\ones^T\brho(\theta)^{-1}\ones)^{-1}\ones^T\brho(\theta)^{-1}\bY\right)^T\\
\hat{\bOmega}_1^* &= (\bX\bGamma_1 - \tY\hat{\bB}_1^T)^T\brho(\theta)^{-1}(\bX\bGamma_1 - \tY\hat{\bB}_1^T)\\
&= \bGamma_1^T\bS_{X|Y}(\theta)\bGamma_1 \\
\hat{\bOmega}_0 &=  (\bX\bGamma_0 - \ones\hat{\mu}_0^T)^T\brho(\theta)^{-1}(\bX\bGamma_0 - \ones\hat{\mu}_0^T) \\
&= \bGamma_0^T\bS_X(\theta)\bGamma_0 \\
\hat{\bSigma}_Y &= (\bY - \ones\hat{\mu}_Y^T)^T\brho(\theta)^{-1}(\bY - \ones\hat{\mu}_Y^T)\\
&= \bS_Y(\theta),
\end{align*}
where $\bS_{X|Y}(\theta)$, $\bS_X(\theta)$ and $\bS_Y(\theta)$ are the sample covariance matrices defined in Equations \ref{samplecov}. Substituting these quantities into the log-likelihood gives the partially maximized log-likelihood
\begin{align*}
L_1(\bX,\bY) =& -\frac{n}{2}\log|\bGamma_1^T\bS_{X|Y}(\theta)\bGamma_1|  -\frac{n}{2}\log|\bGamma_0^T\bS_{X}(\theta)\bGamma_0| -\frac{n}{2}\log|\bS_{Y}(\theta)|  - \frac{p+r}{2}\log|\brho(\theta)|\\
&-\frac{n(p+r)}{2}\log(2\pi + 1). 
\end{align*}
To arrive at objective function \ref{SPEobj}, make the substitution 
\[\log|\bGamma_0^T\bS_X(\theta)\bGamma_0| = \log|\bGamma_1^T\bS_X(\theta)^{-1}\bGamma_1| + \log|\bS_X(\theta)|,\] (see \cite[Lemma~A.13]{cookbook}):
\begin{align*}
L_1(\bX,\bY) =& -\frac{n}{2}\log|\bGamma_1^T\bS_{X|Y}(\theta)\bGamma_1|  -\frac{n}{2}\log|\bGamma_1^T\bS_{X}(\theta)^{-1}\bGamma_1| \\
&-\frac{n}{2}\log|\bS_{X}(\theta)| -\frac{n}{2}\log|\bS_{Y}(\theta)|  - \frac{p+r}{2}\log|\brho(\theta)|-\frac{n(p+r)}{2}\log(2\pi + 1). 
\end{align*}
Still left to be estimated are the regression coordinates $\boldeta$. Here, reverse the conditioning $L(\bX,\bY) = L(\bY|\bX) + L(\bX)$. Parameter $\boldeta$ appears only in $L(\bY|\bX)$, so the second term can be dropped. Differentiating with respect to $\boldeta$ and equating to zero gives $\hat{\boldeta} = \hat{\bGamma}_1^T\hat{\bbeta}_\text{GLS*}$, where GLS coeffecients $\hat{\bbeta}_\text{GLS*}$ are given in Equations \ref{otherparams}.

\subsection{Potential Gains}\label{pga}
The results \ref{glsvar} and \ref{envvar} are proven here. For convenience of notation, we temporarily assume the intercept term is zero in the regression $\vY(s)|\vX(s)$ and the spatial parameter $\theta$ is dropped from $\brho(\theta)$. Recall it is assumed that $\bGamma_1$ and $\theta$ are known.\par
To prove \ref{glsvar}, the law of total variance is first applied
\[\var[\ve(\hat{\bbeta}_\text{GLS})] = \var[\E[\ve(\hat{\bbeta}_\text{GLS})|\bX]] + \E[\var[\ve(\hat{\bbeta}_\text{GLS})|\bX]]. \]
Since $\E[\ve(\hat{\bbeta}_\text{GLS})|\bX] = \bbeta$, the first term is zero. Continuing,
\begin{align*}
\var[\ve(\hat{\bbeta}_\text{GLS})] &= \E[\var[\ve(\hat{\bbeta}_\text{GLS})|\bX]] \\
&= \E[\var[\ve((\bX^T\brho^{-1}\bX)^{-1}\bX^T\brho^{-1}\bY)|\bX]]\\
& = \E[\var[\left(\bI_r \otimes (\bX^T\brho^{-1}\bX)^{-1}\bX^T\brho^{-1}\right)\ve(\bY)|\bX]]\\
&= \E[\left(\bI_r \otimes (\bX^T\brho^{-1}\bX)^{-1}\bX^T\brho^{-1}\right)(\bSigma_{Y|X}\otimes\brho)\left(\bI_r \otimes (\bX^T\brho^{-1}\bX)^{-1}\bX^T\brho^{-1}\right)^T]\\
&= \E[\bSigma_{Y|X} \otimes  (\bX^T\brho^{-1}\bX)^{-1}].
\end{align*}
Since $\var[\ve(\bX)] = \bSigma_X \otimes \brho$, matrix $\bX^* = \brho^{-1/2}\bX$ is equivalent to a matrix of $n$ independent observations of a multivariate normal distribution with variance $\bSigma_X$. Therefore $(\bX^T\brho^{-1}\bX)^{-1} = ({\bX^*}^T\bX^*)^{-1}$ follows an Inverse-Wishart distribtion with expected value $\frac{1}{n-p-2}\bSigma_X^{-1}$, proving \ref{glsvar}.\par
To prove \ref{envvar}, first note that $\boldeta$ is the regression coeffecients for $\vY(s)|\bGamma_1^T\vX(s)$ and $\hat{\boldeta} =  (\bGamma_1^T\bX^T\brho^{-1}\bX\bGamma_1)^{-1}\bGamma_1^T\bX^T\brho^{-1}\bY$. Matrix $\brho^{-1/2}\bX\bGamma_1$ is equivalent to a matrix of $n$ indpendent observations of a multivariate normal distribution with variance $\bOmega_1$. Therefore, by the same logic as above
\begin{align*} 
\var[\ve(\hat{\boldeta})] &= \left(\frac{1}{n-u-2}\right)\bSigma_{Y|X} \otimes  \bOmega_1, 
\intertext{and therefore}
\var[\ve(\hat{\beta})] &= \var[\ve(\bGamma_1\hat{\boldeta})] \\
&= \left(\frac{1}{n-u-2}\right)\bSigma_{Y|X} \otimes  \bGamma_1\bOmega_1\bGamma_1^T,
\end{align*}
proving \ref{envvar}.

\subsection{Asymptotic Distributions}\label{avara}

\subsubsection*{Proof of Theorem \ref{theorem:avar}}

Let $\bE_p$ be the $p^2\times p(p+1)/2$ expansion matrix (sometimes called the duplication matrix) such that for any symmetric $p\times p$ matrix $\bM$, we have $\bE_p\vech(\bM) = \ve(\bM)$. Similarly, let $\bC_p$ be the $p(p+1)/2\times p^2$ contraction matrix (elimination matrix) such that $\bC_p\ve(\bM) = \vech(\bM)$.\par
Assume spatial parameters $\theta$ are known. Let $L$ be the log-likelihood of the full spatial model with no dimension reduction defined by Equations \ref{fullmodel0} and $\bZ$ be the $n\times k$ observation matrix for $\vZ(s)$, where $k$ is the dimension of $\vY(s)$ plus the dimension of $\vX(s)$:
\[L(\bSigma_Z, \mu_Z| \bZ) = -\frac{n}{2}\log|\bSigma_Z|  -\frac{k}{2}\log|\brho_\theta| -\frac{1}{2}\ve(\bZ - \ones\vmu_Z^T)^T(\bSigma_Z \otimes \brho_\theta)^{-1}\ve(\bZ - \ones\vmu_Z^T).\]
With some linear algebra, this can be rewritten as
\[L(\bSigma_Z, \mu_Z| \bZ) = -\frac{n}{2}\log|\bSigma_Z|  -\frac{k}{2}\log|\brho_\theta| -\frac{1}{2}\text{trace}[\bR^T\brho_\theta^{-1}\bR\bSigma_Z^{-1}].\]
where $\bR = \bZ - \ones_n\vmu_Z^T$ is the centered observation matrix. Let $\bR_\theta = \brho_\theta^{-1/2}\bR$ and $\bS_\theta = \frac{1}{n}\bR_\theta^T\bR_\theta = \frac{1}{n}\bR^T\brho_\theta^{-1}\bR$. Rewriting the log-likelihood with this replacement gives
\[L(\bSigma_Z, \mu_Z| \bZ) = -\frac{n}{2}\log|\bSigma_Z|  -\frac{k}{2}\log|\brho_\theta| -\frac{n}{2}\text{trace}[\bS_\theta\bSigma_Z^{-1}].\]
Note that
\[\E[\bR_\theta] = \brho_\theta^{-1/2}\E[\bR] = \bzero,\]
and that
\begin{align*}
\var[\ve(\bR_\theta)] &= \var[(\bI_k \otimes \brho_\theta^{-1/2})\ve(\bR)] \\
&= (\bI_k \otimes \brho_\theta^{-1/2})(\bSigma_Z \otimes\brho_\theta)(\bI_k \otimes \brho_\theta^{-1/2})^T \\
&= (\bSigma_Z\otimes\bI_n).
\end{align*}
Therefore the $n$ rows of $\bR_\theta$ are equivalent to $n$ independent observations of the multivariate normal distribution $N(\bzero, \bSigma_Z)$ and $\E[\bS_\theta] = \bSigma_Z$. The term $-\frac{k}{2}\log|\brho_\theta|$ in the log-likelihood disappears upon differentiation with respect to $\vech(\bSigma_Z)$, leading to the same information matrix for $\vech(\bSigma_Z)$ as the multivariate normal distribution with independent observations, specifically
\[\bJ_{\bSigma} = \frac{1}{2}\bE_k^T(\bSigma_Z^{-1}\otimes\bSigma_Z^{-1})\bE_k.\]
The parameters of $h(\psi)$ are functions of $\bSigma_Z$ and have the same information matrix $\bJ_F$ as in \cite{cookpredictor}. Additionally, the mapping $\psi\rightarrow h(\psi)$ is the same as in \cite{cookpredictor}. Therefore the derivation of the asymptotic distributions follows \cite{cookpredictor} exactly, yielding Theorem \ref{theorem:avar}.

\subsubsection*{Proof of Theorem \ref{theorem:efficient}}

\begin{align}
\bJ_F^{-1} - \bJ_\text{SPE}^{-1} &= \bJ_F^{-1/2}\left(\bI - \bJ_F^{1/2}\bH(\bH^T\bJ_F\bH)^\dagger\bH^T\bJ_F^{1/2}\right)\bJ_F^{-1/2} \\
&=  \bJ_F^{-1/2}\bP_*\bJ_F^{-1/2} \geq 0.
\end{align}
Matrix $\bP_*$ is the orthogonal projection onto the orthogonal complement of the column space of $\bJ^{1/2}\bH$. Since $\bP^* = \bP^*\bP^*$, we have $|\bP^*| = |\bP^*|^2$ implying $|\bP^*|\geq 0$. Matrix $\bP^*$ is therefore semi-positive definite, making $\bJ_F^{-1/2}\bP_*\bJ_F^{-1/2}$ semi-positive definite. This proves the envelope estimator is asymptotically efficient versus the full GLS estimator.

\subsection{Simulation Results}\label{sra}

\begin{figure}[!htbp]
	\centering
	\begin{subfigure}[b]{0.3\textwidth}
		\centering
		\includegraphics[width = \textwidth, keepaspectratio]{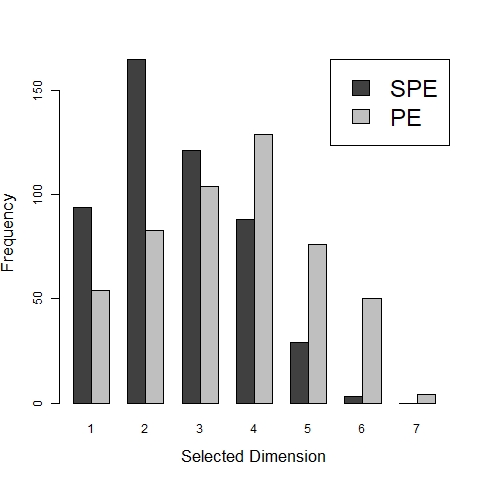}
		\caption{$n = 50$}
	\end{subfigure}
	\hfill
	\begin{subfigure}[b]{0.3\textwidth}
		\centering
		\includegraphics[width = \textwidth, keepaspectratio]{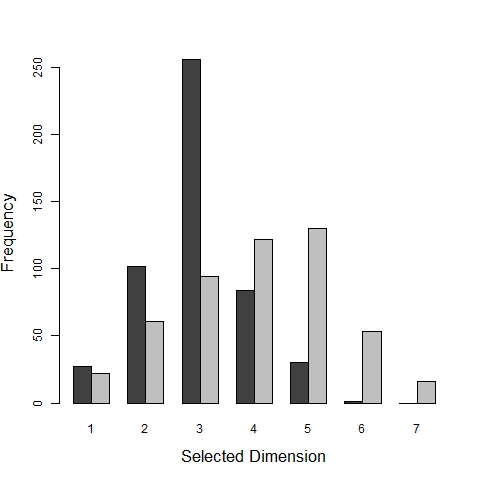}
		\caption{$n = 100$}
	\end{subfigure}
	\hfill
	\begin{subfigure}[b]{0.3\textwidth}
		\centering
		\includegraphics[width = \textwidth, keepaspectratio]{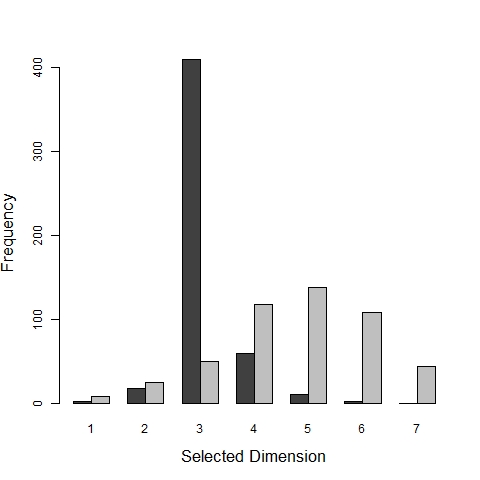}
		\caption{$n = 200$}
	\end{subfigure}
	\hfill
	\caption{Histograms of estimated envelope dimensions for Simulation 2: $u < p$ and $\|\bOmega_0\| \ll \|\bOmega_1\|$. }
	\label{fig:sim2}
\end{figure}

\begin{figure}[!htbp]
	\centering
	\begin{subfigure}[b]{0.3\textwidth}
		\centering
		\includegraphics[width = \textwidth, keepaspectratio]{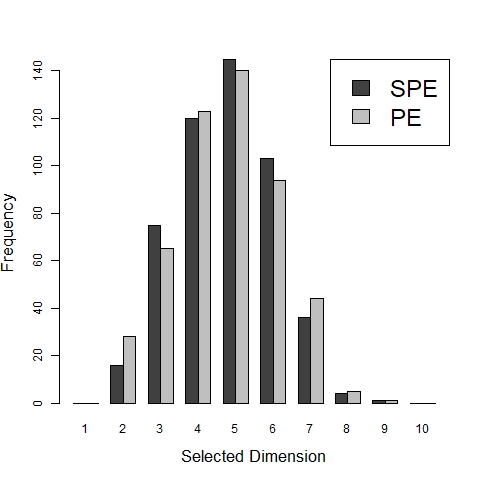}
		\caption{$n = 50$}
	\end{subfigure}
	\hfill
	\begin{subfigure}[b]{0.3\textwidth}
		\centering
		\includegraphics[width = \textwidth, keepaspectratio]{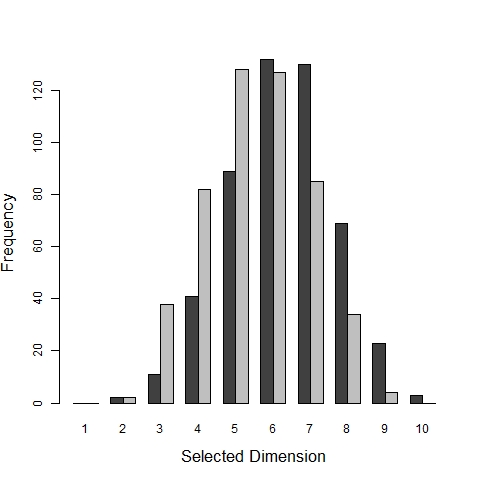}
		\caption{$n = 100$}
	\end{subfigure}
	\hfill
	\begin{subfigure}[b]{0.3\textwidth}
		\centering
		\includegraphics[width = \textwidth, keepaspectratio]{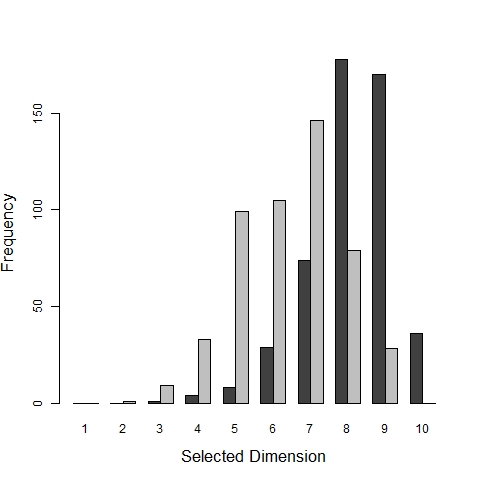}
		\caption{$n = 200$}
	\end{subfigure}
	\hfill
	\caption{Histograms of estimated envelope dimensions for Simulation 3: $u = p$. }
	\label{fig:sim3}
\end{figure}

\begin{figure}[!htbp]
	\centering
	\begin{subfigure}[b]{0.3\textwidth}
		\centering
		\includegraphics[width = \textwidth, keepaspectratio]{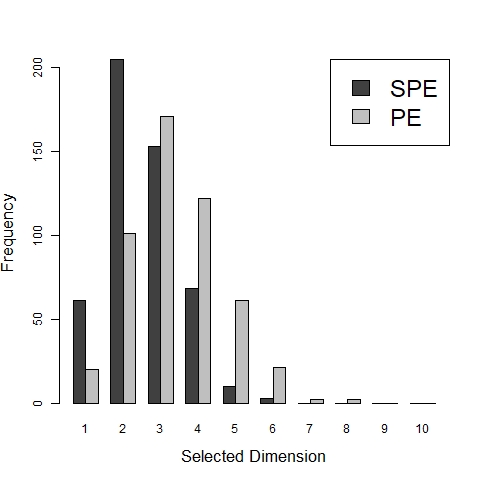}
		\caption{$n = 50$}
	\end{subfigure}
	\hfill
	\begin{subfigure}[b]{0.3\textwidth}
		\centering
		\includegraphics[width = \textwidth, keepaspectratio]{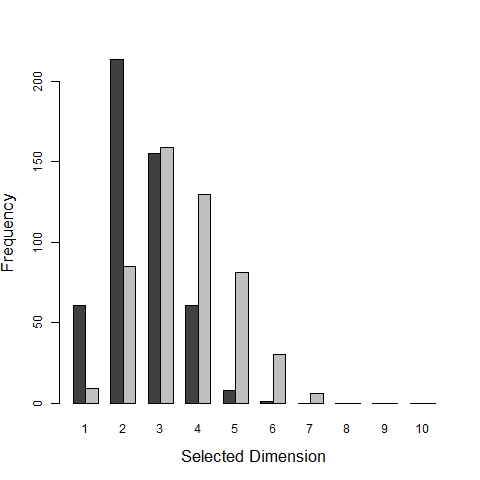}
		\caption{$n = 100$}
	\end{subfigure}
	\hfill
	\begin{subfigure}[b]{0.3\textwidth}
		\centering
		\includegraphics[width = \textwidth, keepaspectratio]{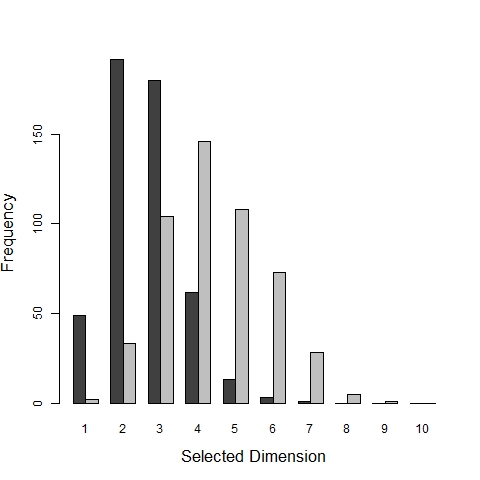}
		\caption{$n = 200$}
	\end{subfigure}
	\hfill
	\caption{Histograms of estimated envelope dimensions for Simulation 4: $u < p$ and $\|\bOmega_1\| \approx \|\bOmega_0\|$. }
	\label{fig:sim4}
\end{figure}

\subsection{Data Analysis}\label{dataa}

Traditional multivariate techniques cannot be directly applied to compositional data because of the variable constraints: all variables for a given observation sum to unity. Thus, it has become common to analyze the log-ratios of the components. Let $\vZ = [Z_1, \ldots, Z_k]$ be a $k$-dimensional composition (or subcomposition). Then the $k-1$ dimensional transformed variable
\[\vX = [\log(Z_1/Z_k)\ \log(Z_2/Z_k)\ \cdots\ \log(Z_{k-1}/Z_k)]^T, \]
is unconstrained. Furthermore, if the composition $\vZ$ is assumed to follow a logistic normal distribution, then $\vX$ follows a multivariate normal distribution. The choice of denominator in the ratio, given above as the last component of the composition, is arbitrary, and equivalent inference is obtained for any choice of denominator \citep[Chapter~5]{aitchison}.\par

For the data analysis in this paper, the predictors were chosen to be the relative log-concentrations of the major chemical elements. Let $\text{MCE}_1, \ldots, \text{MCE}_{11}$ represent the eleven major chemical elements. The major chemical elements are examined as a subcomposition, with 
\[Z_j = \text{MCE}_j/(\sum_{w=1}^{11}\text{MCE}_w),\] 
normalizing so that the relative proportions of the major chemical elements are studied. The major chemical elements are normalized as a separate subcompositon in order to remove any trivial, compositional relationships between the predictors and the response, for example, an increase in potassium oxide concentration causes a decrease in REE concentration simply by leaving less room in the composition. Rather, the goal is to analyze the effect of the relative proportions of major chemical elements on the REE concentration, for example, in increase in potassium oxide relative to the other major chemical elements makes higher REE concentrations likely. To remove the constraints on the predictors, a log-ratio transformation is applied
\[X_j = \log(Z_j/Z_{11}),\ j=1,\ldots,10\]
leaving $p = 10$ predictor variables, with major chemical element aluminum oxide (Al2O3) arbitrarily chosen as the denominator.\par 
The REE element concentrations are represented in parts per million, between $0$ and $10^6$. The regression response is chosen to be the log-concentration of REEs relative to all other parts of the composition
\[Y = \log(REE/(10^6 - REE)). \]
a monotone increasing transformation that gives a response that is unconstrained, interpretable and congruent with the goals of the analysis. \par
Some of the observations had chemical element concentrations below the detection threshold of the instrument used to analyze the sample. Specifically, three samples had a major chemical element titanium dioxide (TiO2) below the detection threshold and ten samples had at least one rare earth element below the detection threshold. Rather than omit these samples, the values were replaced with half of the detection threshold. Because so few samples had a major chemical element below the detection threshold, and any REE below the detection threshold will have little impact on the total REE concentration, this choice of replacement had little impact when compared to any other reasonable replacement scheme.

\end{document}